\documentclass[twocolumn,letterpaper,superscriptaddress,prl]{revtex4}
\usepackage{graphicx}
\usepackage[usenames]{color}
\usepackage{multirow}
\usepackage{verbatim}
\usepackage{amsmath}
\usepackage{amssymb}
\usepackage{mathrsfs}
\usepackage{url}
\usepackage{float}

\begin{document}

\title{Quantitative Current-Voltage Characteristics in Molecular Junctions from First Principles}
\author{Pierre~Darancet}
\email{pdarancet@lbl.gov}
\affiliation{Molecular Foundry, Lawrence Berkeley National Laboratory, Berkeley, CA, USA}

\author{Jonathan~R.~Widawsky}
\affiliation{Department of Applied Physics and Applied Mathematics, Columbia University, New York, NY, USA}

\author{Hyoung~Joon~Choi}
\affiliation{Department of Physics, Yonsei University, Seoul, South Korea}

\author{Latha~Venkataraman}
\email{lv2117@columbia.edu}
\affiliation{Department of Applied Physics and Applied Mathematics, Columbia University, New York, NY, USA}

\author{Jeffrey~B.~Neaton}
\email{jbneaton@lbl.gov}
\affiliation{Molecular Foundry, Lawrence Berkeley National Laboratory, Berkeley, CA, USA}
\email{jbneaton@lbl.gov}

\begin{abstract}
Using self-energy-corrected density functional theory (DFT) and a coherent scattering-state approach, we explain  current-voltage (IV) measurements of four pyridine-Au and amine-Au linked molecular junctions with quantitative accuracy. Parameter-free many-electron self-energy corrections to DFT Kohn-Sham eigenvalues are demonstrated to lead to excellent agreement with experiments at finite bias, improving upon order-of-magnitude errors in currents obtained with standard DFT approaches. We further propose an approximate route for prediction of quantitative IV characteristics for both symmetric and asymmetric molecular junctions based on linear response theory and knowledge of the Stark shifts of junction resonance energies. Our work demonstrates that a quantitative, computationally inexpensive description of coherent transport in molecular junctions is readily achievable, enabling new understanding and control of charge transport properties of molecular-scale interfaces at large bias voltages. 
\end{abstract}

\maketitle

There is significant interest in using organic molecules as active components in next-generation energy conversion devices such as organic photovoltaics~\cite{OPV} and dye-sensitized solar cells~\cite{Dye}, where a roadblock to higher efficiencies is quantitative understanding and control of charge transport phenomena at interfaces. Understanding electronic energy level alignment and transport at interfaces between active organic layers and conducting electrodes has been particularly challenging~\cite{MetalMolInterfaces1,MetalMolInterfaces2}. However, recent scanning tunneling microscope-based break-junction (STM-BJ) experiments~\cite{Tao,VenkataramanNature}
 of molecular junctions --devices formed by trapping organic molecules between macroscopic metallic electrodes-- have reported robust conductance~\cite{Tao,VenkataramanNature,BreakJunctionExperiments1,BreakJunctionExperiments2,BreakJunctionExperiments3},   thermopower~\cite{Thermopower1,Thermopower2,ThermopowerWidawsky},   switching behavior~\cite{Switching1,Switching2,SuYingNatNano}, 
 quantum interference effects~\cite{Mayor,Interference1,Interference2},   spin-filtering phenomena~\cite{SpinFiltering1,SpinFiltering2,SpinFiltering3},   and even full nonlinear IV characteristics~\cite{Mayor,Widawsky,ExpIV},  establishing such junctions as unique and revealing windows into the physics of charge transport at the molecular scale.

Given the diversity of known synthesizable organic molecules, and a lack of intuition connecting transport phenomena to a specific molecule and interface, a rigorous yet pragmatic quantitative theory capable of predicting charge transport properties of molecular junctions with chemical specificity is needed. Charge transport calculations require solution of the electronic structure for an open system out of equilibrium~\cite{Reviews1,Reviews2,Reviews3,Reviews4,Reviews5,diventra}. The majority of prior theoretical studies have relied on a Landauer approach, simplified to treat electronic interactions at a mean-field level within density functional theory (DFT) using either Green's functions~\cite{NEGFimplementation1,NEGFimplementation2,NEGFimplementation3,NEGFimplementation4} or scattering-states~\cite{Scarlet}. However, within common approximations to DFT, such as generalized gradient approximations (GGAs) and hybrid functionals, Kohn-Sham orbital energies are known to yield unsatisfactory level alignment~\cite{dellangela, Tamblyn, Rignanese,SuYingPRL,KSWrongLevel1,KSWrongLevel2}, leading to overestimated conductance~\cite{Reviews1,Reviews2,Reviews3,Reviews4,Reviews5,diventra} and even incorrect trends~\cite{quekACSNano} relative to experiment. 

Improved treatment of exchange and correlation effects within the junction~\cite{quek1,quek2,quek3,diventra,quekACSNano,Rignanese,SuYingPRL,Theorytransport1,Theorytransport2,Theorytransport3,Theorytransport4,Theorytransport5,Theorytransport6} has been shown to lead to better agreement with experiment. In particular, recent calculations based on many-body perturbation theory within the GW approximation~\cite{quek1,quek2,quek3,NeatonPRL, SuYingNatNano} have been demonstrated to amend junction level alignment and have resulted in conductance in quantitative agreement with measured values~\cite{quek1,quek2,quek3, SuYingNatNano}. 
However, these approaches rely on computationally expensive methods~\cite{Rignanese,SuYingPRL,Theorytransport1,Theorytransport2,Theorytransport3,Theorytransport4,Theorytransport5,Theorytransport6}, and their accuracy come at the cost of their tractability for junctions relevant to experiments; moreover,  direct comparisons to experiment have been limited to low-bias measurements in the linear response regime. 
The ability to measure, reliably, conductance and nonlinear IV characteristics, and the promise of exploring new phenomena at higher bias, calls for quantitative computational studies of truly nonequilibrium steady-states.

In this Letter, we demonstrate and apply a quantitative framework capable of explaining measured IV characteristics for four molecules  --4,4' diamino-stilbene (DAS), bis-(4-aminophenyl)acetylene (APA), 1,6-hexanediamine (HDA), and 4,4'-bipyridine (BP)-- in contact with gold electrodes for biases as high as 1V. Whereas the magnitude of the currents and shape of the measured IV characteristic are not reproduced by a standard finite-bias DFT approach~\cite{NEGFimplementation1,NEGFimplementation2,NEGFimplementation3,NEGFimplementation4,Scarlet}, a method based on the GW approximation~\cite{NeatonPRL}, successful for low-bias conductance and extended here to finite bias, leads to excellent agreement with experiment. Our approach enables inexpensive and accurate calculations of coherent transport properties, which can be used for the design of functional molecular junctions with nonlinear IV characteristics.
		
\begin{figure*}
      \includegraphics[height=57mm, width=160mm]{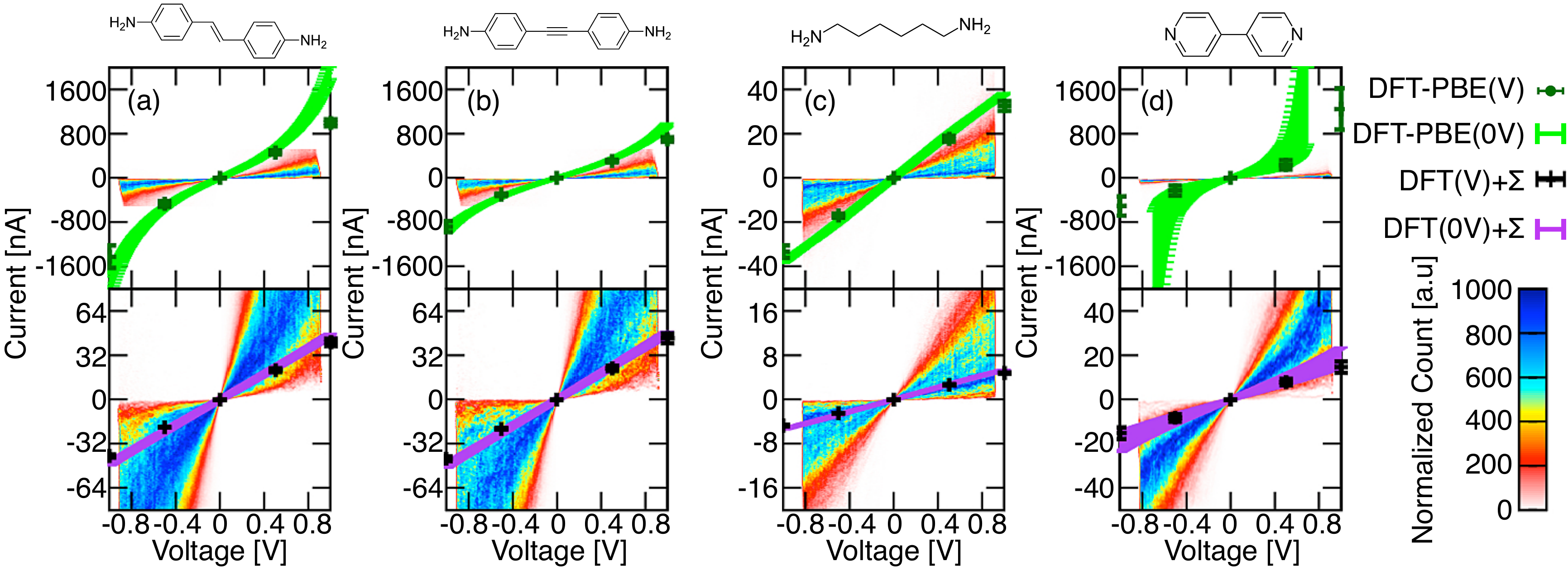}
      \caption{ $I\left(V\right)$ characteristics for (a) DAS, (b) APA, (c) HDA, and (d) BP junctions. 
      Top:  Experiment (color map), DFT-PBE (green line), and DFT-PBE(V) (dark-green points). 
       Bottom: Experiment (color map), DFT$+\Sigma$ (grey line), and DFT(V)$+\Sigma$ (black points). 
       Error bars are added to the computed currents to indicate the spread associated with the three different contact geometries used here (see SI).  
       Measured IV characteristics for DAS, APA, and HDA junctions are adapted from~\cite{Widawsky}. Lewis structures of each molecule at top. 
       }
     \label{Fig:Current}
\end{figure*}

Previous studies~\cite{quek1,quek2,quek3,SuYingNatNano} have shown amine- and pyridine-bonded molecules can bind preferentially to undercoordinated gold atoms. Building on those studies, we construct three geometries for each junction with trimer (3 gold atoms), trimer-adatom, and adatom binding motifs (see SI for geometries). We relax all junctions using DFT within the GGA of Perdew, Burke, and Ernzenhof (PBE)~\cite{PBE} and a double-$\zeta$-basis set as implemented in SIESTA~\cite{Siesta1,Siesta2}. Details of our DFT calculations are provided in previous work~\cite{quek1,quek2,quek3}. Atomic positions are relaxed until Hellmann-Feynman forces are smaller than $0.04$~eV/\AA. We model our system by $7$-layers of $16$ gold atoms on both sides, the last $4$ layers being constrained to the bulk geometry. Initial trial geometries are adapted from previous works~\cite{quek1,quek2,quek3,SuYingNatNano,Alkane1,Alkane2}. We consider BP junctions in a ``high conductance'' configuration (HighG), following the notation of Ref.~\cite{SuYingNatNano}.
 
In what follows, the conductance and IV characteristics are computed from a Landauer-like formula using a coherent scattering-state approach, and include exchange and correlation contributions that correct for zero-bias junction level alignment, as explained below. In this framework, the IV characteristic is expressed in terms of an energy- and bias-dependent transmission function $ \mathcal{T} \left( \omega;V \right)$ as
\begin{eqnarray}
 I\left( V \right) &=&  \frac{2e}{h} \int_{-\infty}^{+\infty}   d\omega \mathcal{T} \left( \omega;V \right) \left[ f\left(\omega+e\frac{V}{2}\right)  -  f\left(\omega -e\frac{V}{2}\right)  \right],
\end{eqnarray}
where $f$ is the lead Fermi-Dirac occupation function and $\omega$ has units of energy. The transmission function is computed as $\mathcal{T} \left( \omega;V \right) = \textrm{tr} \left[ t\left( \omega ;V\right)  t^\dagger\left( \omega ;V\right) \right]$, where $t$ and $ t^\dagger$  are the transmission coefficients of the junction scattering states computed as a function of bias~\cite{Scarlet, NeatonCompPhysComm}. The zero-bias conductance is determined via linear response, as $\mathcal{G}=\left. \frac{dI}{dV} \right|_{V=0} = \frac{2e^2}{h} \mathcal{T} \left( E_F;V=0 \right)$, where $E_F$ is the junction Fermi level at zero bias.

We obtain the self-consistent steady-state density matrix from DFT-PBE as described in~\cite{Scarlet, NeatonCompPhysComm} using an $8\times8$ $k_{//}$-mesh and in a manner equivalent to prior work~\cite{Scarlet,NEGFimplementation1,NEGFimplementation2,NEGFimplementation3,NEGFimplementation4}. For finite bias calculations, referred to as DFT(V), the density matrix includes a real-axis integration of the scattering-states in the bias window on an adaptive energy grid ~\cite{Scarlet}  with a resolution of up to $10^{-7}$eV in the vicinity of the molecular resonances. The transmission function is generated in a subsequent step using a $16\times16 k_{//}$-mesh. 
                   
To correct for inaccuracies associated with DFT-PBE Kohn-Sham eigenvalues for quasiparticle energy level alignment, we employ a physically-motivated electron self-energy correction to the molecular orbital energies in the junction, DFT$+\Sigma$, following Ref.~\cite{quek1,quek2,quek3}. Formally derivable in the weak coupling limit of the $GW$ approximation~\cite{GW,Inkson,quek3}, this adjustable-parameter-free model self-energy acts on the molecular subspace and consists of two terms: $i)$ a gas-phase correction accounting for the difference between DFT highest occupied molecular orbital (HOMO) and lowest unoccupied molecular orbital (LUMO) energies and their gas-phase ionization potential (IP) and electron affinity (EA) ($\Sigma_{\textrm{GP}}$); and $ii)$ an "image charge" term, accounting for the polarization energy associated with static non-local correlations between the electrons (or holes) on the molecule and in the metal that close the gap of the molecule upon absorption ($\Sigma_{\textrm{CORR}}$). In this work, we calculate $\Sigma_{\textrm{GP}}$ using a $\Delta$SCF method, and following previous studies~\cite{NeatonPRL,quek1,quek2,quek3,Tamblyn}, we approximate $\Sigma_{\textrm{CORR}}$ with an electrostatic image charge model (see SI for details). Since these self-energy corrections are large compared to those for the metallic bulk and surface Au states~\cite{RangelGold}, especially for states near E$_F$, we neglect similar corrections to the Au states. Values for $\Sigma_{\textrm{GP}}$ and $\Sigma_{\textrm{CORR}}$ for each junction appear in the SI. This approach was shown to lead to quantitative agreement with photoemission  experiments regarding the level alignment of benzene diamine derivatives on flat gold (111) \cite{dellangela}.

In \ref{Tab:0bias}, we report computed zero-bias conductances for  DAS, APA, HDA, and BP junctions for both DFT-PBE and DFT$+\Sigma$, and compare with experiments~\cite{SuYingNatNano,Widawsky}. As with previous work~\cite{quek1,quek2,quek3,SuYingNatNano,quekACSNano,ThermopowerWidawsky}, we find DFT-PBE overestimates measured low-bias conductances for DAS, APA, and BP junctions by more than an order of magnitude. An exception is the HDA junction, where the overestimate is just a factor of 4, consistent with Ref.~\cite{Alkane1,Alkane2}. In contrast, DFT$+\Sigma$ improves agreement with experiment significantly, predicting low-bias conductances to within a factor of two and shifting frontier orbital resonances to higher energies of $1.5$~eV (BP-LUMO)  or more away from $E_F$. For DAS, APA, and BP junctions, the total correction $\Sigma_{\textrm{GP}} +\Sigma_{\textrm{CORR}}$ opens the gap between junction HOMO and LUMO molecular resonances by more than $2.5$~eV. For HDA, we find that $\Sigma_{\textrm{GP}}$ and $\Sigma_{\textrm{CORR}}$ are of the same magnitude but opposite sign, resulting in a modest net correction of $0.15$~eV, explaining its relatively accurate DFT-PBE zero-bias conductance by a cancellation of errors.

\begin{table}
\begin{tabular}{ l*{4}{c}r }
 Junction & Exp   & DFT-PBE & DFT+$\Sigma$     \\
\hline
 DAS  & $10$   & $108-125$ & $5.7-5.9$  \\
APA  & $8$  & $75-83$ &  $6.0-6.4$ \\
 HDA  &  $1.2$  &$5.1-5.4$ & $0.7$ \\
 BP    & $6$   & $22-94$ & $1.5-2.8$ 
\end{tabular}
   \caption{Measured and calculated zero-bias differential conductances (in units of $10^{-4}G_0$) for the different molecule-gold junctions under study. The computed spread reflects the lowest and highest conductance of the three geometries considered for each junction (see SI for details). Experimental data for BP comes from Ref.~\cite{SuYingNatNano}. All other experimental data is taken from Ref.~\cite{Widawsky}.}
  \label{Tab:0bias}
\end{table}

Calculated IV characteristics for each junction are shown in \ref{Fig:Current} and compared with experiments. We first generate a DFT-PBE steady-state charge density, including a real-axis integration of the scattering-states in the bias window, at each bias voltage; the current is then determined from an integration of $\mathcal{T} \left( \omega;V \right)$, subsequently computed on a dense energy grid, as described above. Then, in one approach, which we refer to as DFT-PBE(V), $\mathcal{T} \left( \omega;V \right)$ is generated from DFT-PBE junction electronic structure, which is known to overestimate the zero-bias conductance; in the other approach, referred to as DFT(V)$+\Sigma$, $\mathcal{T}\left( \omega;V \right)$ is generated from DFT$+\Sigma$ junction electronic structure, with $\Sigma$ determined at zero-bias.

As can be seen in \ref{Fig:Current}, DFT(V)$+\Sigma$ leads to excellent quantitative agreement with experiment, whereas the DFT-PBE(V) currents are too large by an order of magnitude (or more). Interestingly, integration of the zero-bias DFT-PBE transmission function (gray curves), assuming a uniform potential drop across the junction, results in a good estimate of the DFT-PBE(V) IV characteristics for DAS, APA, and HDA junctions, suggesting that $\mathcal{T}\left( \omega;V \right)$ is weakly dependent on applied bias for these junctions. A similar integration results in an overestimate by a factor of $8$ to $10$ for the BP junctions, which we explain below.  

\begin{figure*}
      \includegraphics[height=56mm, width=160mm]{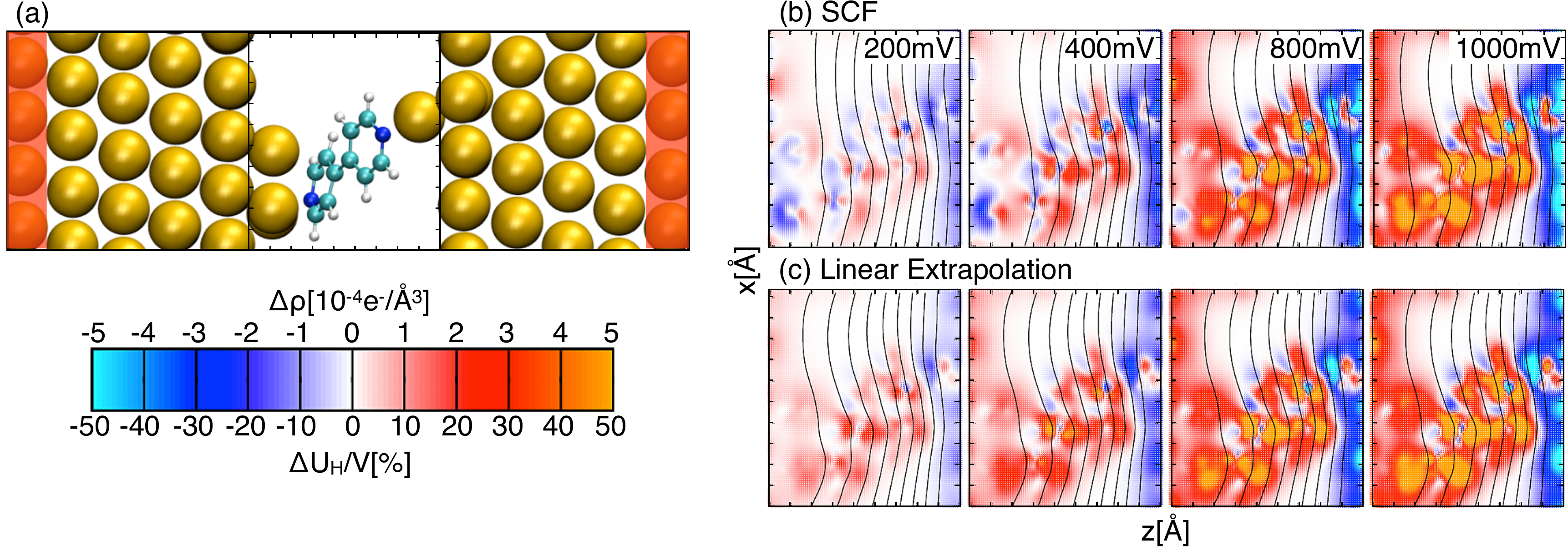}
      \caption{
      Junction structure, and DFT-PBE differences in electronic density and Hartree potentials in a BP junction for biases up to $1000$ mV. 
      (a) Junction structure. The red shaded areas indicate where densities and potentials are fixed to their bulk values;  the frame in (a) indicates the region plotted in (b) and (c).
      (b) $\Delta \rho$ and $\Delta U_H$, determined via self-consistent DFT-PBE(V) calculations. 
      (c) Linear extrapolation result using  $\rho$ and $U_H$ at zero bias and $\frac{\partial \rho}{\partial V} $ and $ \frac{\partial U_H}{\partial V}$ calculated at $-100$~mV. 
      The color map represents differences in density, and the contour lines indicate isovalues of  the change in Hartree potential scaled by the applied bias, i. e. $\frac{\Delta U_H}{V} = -50\%, -40\%\dots +50\%$.  
      }
     \label{Fig:Density}
\end{figure*}

In \ref{Fig:Density}, we plot the evolution of the DFT-PBE charge density, $\Delta\rho$, and electrostatic Hartree potential, $\Delta U_H$, with bias. $\Delta\rho \left(V\right)$ and $\Delta U_H\left(V\right)$ are both computed relative to zero-bias densities and potentials, respectively.  We compare this evolution to that generated assuming linear responses of the density and the Hartree potential, \textit{i.e.}  $\Delta\rho \sim \left. \frac{\partial \rho}{\partial V}\right|_{V=0} \times V$ and $\Delta U_{H} \sim \left. \frac{\partial U_H}{\partial V} \right|_{V=0} \times V$, respectively. We find that all $4$ junctions remain in the linear response regime for the experimental range of bias ($\pm 0.8$ V). This finding is consistent with the fact that the DFT+$\Sigma$ molecular resonances are far from the bias window in these junctions, even for the highest biases achieved, leading to modest changes in occupation numbers. Indeed,  a simple estimate of the change in charge density under bias, obtained by fitting the DFT(V=0)$+\Sigma$ transmission function to a symmetric Lorentzian model for all four junctions (see SI), leads to a maximum expected change in occupation by only $0.007$ $e^-$.

Interestingly for BP junctions (\ref{Fig:Density}), the DFT-PBE(V) densities and Hartree potentials remain in the linear response regime to $1$~V, even though the DFT-PBE level alignment erroneously places  the LUMO resonance in the bias window at this voltage. This indicates that, despite the incorrect DFT-PBE energy level alignment, self-consistent DFT-PBE(V) results are nonetheless in line with the DFT$+\Sigma$ level alignment for the prediction of the densities under bias and do not show significant charging or any (non-physical) deviation from linear response. 

\begin{figure}
    \includegraphics[height=60mm, width=80mm]{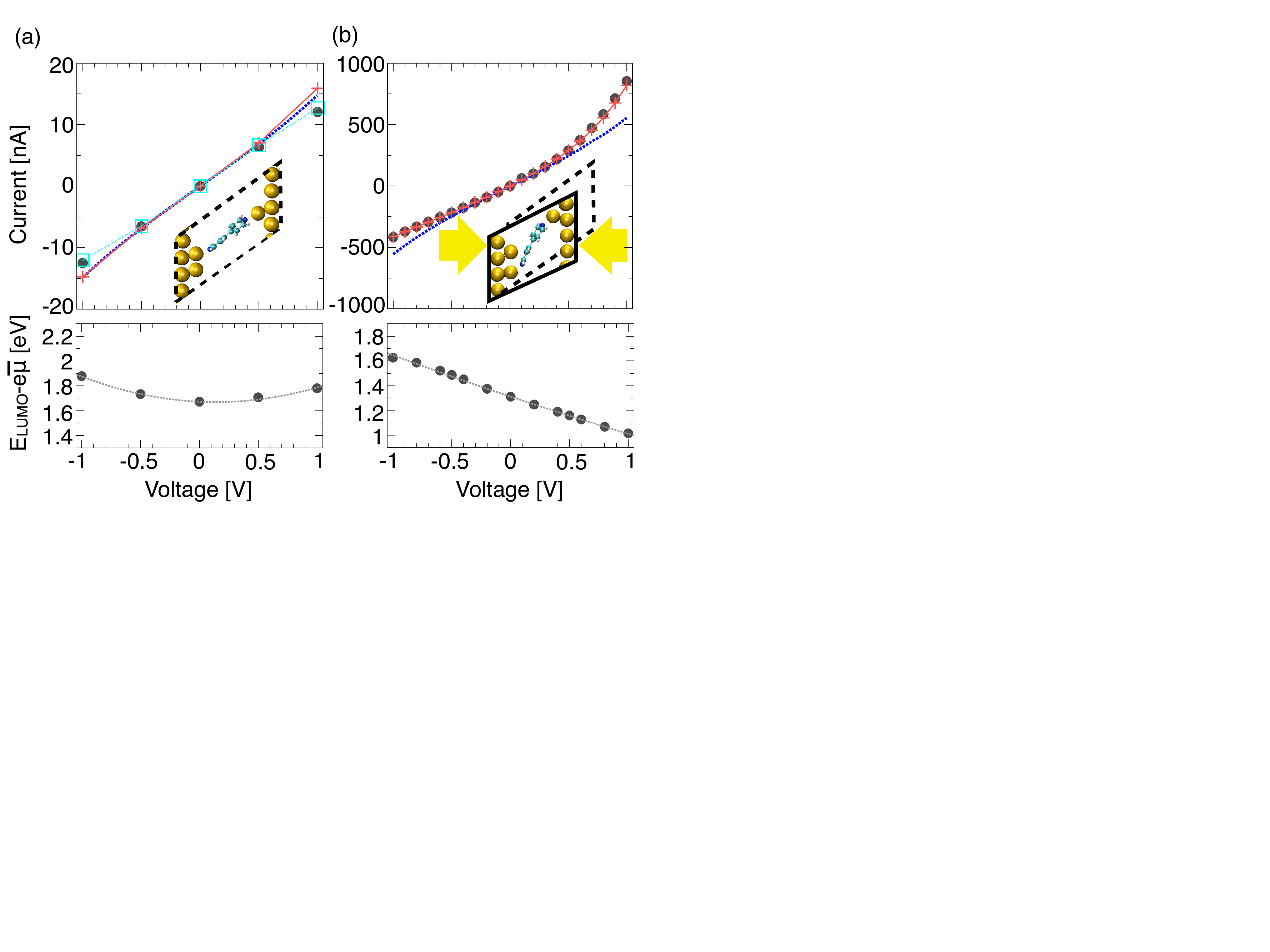}
    \caption{
Molecular resonance energies and IV characteristics for a (a) symmetric and (b) asymmetric BP junction. 
Top: Calculated $I(V)$ using self-consistent DFT(V)$+\Sigma$ (black dots), DFT(V=0)$+\Sigma$ (blue dotted line), and one-shot DFT(V=0)$+\Sigma(V)$ with both first-order (red crosses) and second-order (cyan squares) corrections to the zero bias Hamiltonian.  Insets: Structure of the junctions. 
Bottom: Frontier orbital resonance energies with respect to the average chemical potential of the junction as a function of applied voltage, self-consistently calculated with DFT(V)$+\Sigma$ (dots), and best polynomial fits (line) of the change in resonance energy with $V$, \textit{i.e.} (a) $\Delta E(V)=-0.04V+0.2V^2$ and (b) $\Delta E(V)=-0.32V+0.02V^2$.}
   \label{Fig:Levels}
\end{figure}

In \ref{Fig:Levels}, we report the evolution of the molecular resonances energies with bias for two BP junctions. For a symmetric  junction, the resonance energies vary as  $V^2$, as expected from a quadratic Stark effect, with a maximum shift (relative to the average chemical potential) of less than $0.06$~eV for DAS and APA, and up to  $0.2$~eV  for HDA and BP (\ref{Fig:Levels} (a)). The small magnitude of these shifts explains why a simple integration of the zero-bias $\mathcal{T}\left( \omega;V=0 \right)$ is so effective for these junctions. However, if the junctions are more asymmetric, such as the BP junction shown in \ref{Fig:Levels} (b),  the resonance energies may vary more substantially. In fact, for the asymmetric BP junction shown in \ref{Fig:Levels} (b), the LUMO resonance energy shifts linearly with $V$, with small second-order corrections to the energy. Appreciable first-order corrections to the wavefunction are also evident in this case, via a $50\%$ reduction in the resonant peak maximum. Integration of the zero-bias $\mathcal{T}\left( \omega;V=0 \right)$ is clearly much less effective in this case, as $\mathcal{T}\left( \omega;V \right)$ is varying significantly with $V$. 

Since the Stark shifts of the LUMO resonance for BP junctions are well-described by a simple function of $V$, a modified self-energy correction of the form $\Sigma\left(V\right) = \Sigma_{\textrm{GP}} +\Sigma_{\textrm{CORR}}+ \Delta E \left(V\right)$, with $\Delta E \left(V\right)$ from a polynomial fit of resonance energy shifts from just a few finite bias calculations, results in predicted currents in close agreement with those computed self-consistently at each V. 
Since our ``Stark-corrected''  self-energy $\Sigma\left(V\right)$ is calculable at very low-bias or from the zero-bias polarizability, it can be used for efficient and quantitative prediction of nonlinear IVs and rectification ratios, at least for off-resonance tunneling through non-degenerate levels~\cite{DarancetToBePublished}. 

Our results for BP junctions, shown in Figures~\ref{Fig:Density} and \ref{Fig:Levels}, are consistent with our findings for DAS, APA, and HDA. For each junction, we can draw the following conclusions. First, extrapolations using the zero-bias response functions $\left. \frac{\partial \rho}{\partial V} \right|_{V=0}$ and $\left. \frac{\partial U_H}{ \partial V} \right|_{V=0}$ result in accurate densities and potentials at other experimentally achievable biases, obviating the need in these cases to do laborious self-consistent calculations at many bias voltages. Second, self-consistency at DFT-PBE level is apparently sufficient for these junctions, despite errors in level alignment, provided that the actual resonances are far from the bias window. Third, a straightforward Stark correction $\Delta E \left( V \right)$ to the DFT-PBE resonance peak energy at zero-bias $E \left(V=0\right)$, along with a zero-bias self-energy correction, yields accurate IV characteristics, even for asymmetric junctions. 

The success of DFT$+\Sigma$ at finite-bias is noteworthy and deserves further comment. Because of the small polarizabilities of the molecules considered here, modifications of $\Sigma_{\textrm{GP}}$ in relevant electric fields are just $0.01$~eV for DAS and APA, $0.02$~eV for HDA, and $-0.002$~eV for BP, a mere few  $\%$ correction to the zero-bias self-term~\cite{SEcorrect}. Moreover, using the DFT$+\Sigma$ frontier orbital energies, the weak densities of states in the bias window for all junctions give rise to negligible changes in occupation under bias, smaller than $0.007~e^-$, implying equally negligible changes in $\Sigma_{\textrm{CORR}}$ for these junctions in the off-resonance limit. Thus corrections to $\Sigma\left(V=0\right)$ at finite bias will be extremely small. We expect the good agreement between  the DFT$+\Sigma$ IV characteristics and experiments in \ref{Fig:Current} to deteriorate for more polarizable molecules (due to the changes in the $\Sigma_{\textrm{GP}}$  term), stronger coupling of  resonances to the lead states (via a breakdown of the DFT$+\Sigma$ approximation), and for molecular junctions beyond the linear response regime of the density (associated with changes in $\Sigma_{\textrm{CORR}}$ term, and high-order terms in V in $\Delta E\left(V\right) $), \textit{e.g.} for molecular resonances closer to the Fermi energy of the metal.

To conclude, we have computed IV characteristics for four different molecular junctions, and compared them with STM-BJ experiments. While state-of-the-art methods based on DFT-PBE largely overestimate measured currents, our pragmatic first-principles approach based on one-shot self-energy corrections that improve level alignment in the junction leads to excellent quantitative agreement. This method opens up new avenues for computationally inexpensive and quantitative modeling of non-equilibrium steady-state properties of molecular junctions: our results suggest that the finite-bias transmission function for both symmetric and asymmetric molecular junctions can be well-approximated by calculating the changes in zero-bias resonance energies under small perturbative fields. These findings offer the possibility to more rapidly understand and predict functional nonlinear properties of junctions, such as the rectification, accurately and efficiently.

We thank David Strubbe, Su Ying Quek, Peter Doak, and David Prendergast for discussions. 
Portions of this work were performed at the Molecular Foundry and within the Helios Solar Energy Research Center, and both were supported by the Office of Basic Energy Sciences of the U.S. Department of Energy under Contract No. DE-AC02-05CH11231.  
STM-BJ experiments were supported as part of the Center for Re-Defining Photovoltaic Efficiency Through Molecular-Scale Control, an Energy Frontier Research Center funded by the U.S. Department of Energy (DOE), Office of Science, Office of Basic Energy Sciences under Award DE-SC0001085. 
L.V. thanks the Packard Foundation for support. 
H.J.C. acknowledges support from NRF of Korea (Grant No. 2011-0018306) and KISTI Supercomputing Center (Project No. KSC-2012-C2-14). Computational resources were provided by NERSC. Experimental methods, level alignment and self-energy corrections, analytical changes in the density of a Lorentzian model, basis sets, geometries, plot of the self-consistent density matrices, Hartree potentials and transmission functions at finite biases are reported in the supporting information. This material is available free of charge via the Internet at http://pubs.acs.org.


\begin{thebibliography}{achemso}
\bibitem{SEcorrect}
The GP-term in the latter expression is computed using the $\Delta$-SCF result using B3LYP as implemented in the QChem code~\cite{QChem}. Results for the molecule in the field are calculated using 61 calculations of total energies of the neutral, cationic, anionic species for fields ranging from $-8.3\times10^{-2}$V/\AA  to $+8.3\times10^{-2}$V/\AA. The monopole term is calculated fitting the total energies of the cations and anions on $21$ points for fields ranging  $-8.3\times10^{-4}$V/\AA to $8.3\times10^{-4}$V/\AA, and then subtracted. The non-local part of the static correlations is approximated using an image charge model on the pseudo-charge density of the levels given by SIESTA as described in~\cite{quek1,quek2,quek3}. The position of the image plane is calculated using the procedure of  Lam \& Needs~\cite{LamNeeds}.
\end{thebibliography}
\end{document}